\documentclass[prb,twocolumn,showpacs]{revtex4}

\usepackage{amsmath}
\usepackage{bm}
\usepackage{times}
\usepackage{graphicx}
\usepackage{psfrag}

\begin{document}

\vspace*{0.5ex}

\title{Theory for transport through a single magnetic molecule: Endohedral 
\mbox{\boldmath$\mathrm{N@C_{60}}$}}

\author{Florian Elste}
\email{felste@physik.fu-berlin.de}
\affiliation{Institut f\"ur Theoretische Physik, Freie Universit\"at Berlin,
Arnimallee 14, D-14195 Berlin, Germany}
\author{Carsten Timm}
\email{timm@physik.fu-berlin.de}
\affiliation{Institut f\"ur Theoretische Physik, Freie Universit\"at Berlin,
Arnimallee 14, D-14195 Berlin, Germany}

\date{October 25, 2004}

\begin{abstract}
We consider transport through a single $\mathrm{N@C_{60}}$ molecule, weakly
coupled to metallic leads. Employing a density-matrix formalism we derive
rate equations for the occupation probabilities of many-particle states of
the molecule. We calculate the current-voltage characteristics and the
differential conductance for $\mathrm{N@C_{60}}$ in a break
junction. Our results reveal Coulomb-blockade behavior as well as a fine
structure of the Coulomb-blockade peaks due to the exchange coupling of the
$\mathrm{C_{60}}$ spin to the spin of the encapsulated nitrogen atom.
\end{abstract}

\pacs{	
85.65.+h, 
75.50.Xx 
}

\maketitle

\section{Introduction}

The rapid progress of miniaturization of electronic devices has led to
chip features smaller than $100\,\mathrm{nm}$, for which standard
semiconductor technology reaches its limit. One proposed
solution is a transistor consisting of a single molecule. In recent years
transport through single molecules has been studied quite
extensively\cite{Joachim,EmK02,Nitzan,XuR03,Mitra,Koch}---for example, in break
junctions.\cite{RZM97,Park,Weber}
\emph{Inelastic} transport occurs due to the interaction of electrons with
internal vibrational or magnetic degrees of freedom of the molecules.
Transport through magnetic molecules\cite{Park2,Liang,Durkan} is
particularly interesting also from the point of view of 
\emph{spintronics}---i.e., the idea of exploiting the electron spin in electronic devices. While
most molecules are normally nonmagnetic, there are exceptions such as
endohedral $\mathrm{N@C_{60}}$---i.e., a nitrogen atom encapsulated in a
$\mathrm{C_{60}}$ cage.\cite{Almeida} It is known that the
encapsulated atom retains its $p$ electrons,\cite{Almeida} leading to a
localized spin $S_{\mathrm{N}}=3/2$. There are fascinating ideas of
employing this spin in a quantum computer.\cite{Harneit}

In this paper we propose to measure the current through a single
$\mathrm{N@C_{60}}$ molecule in a break junction and we calculate the 
current-voltage (\textit{I-V}) characteristics and the differential
conductance $dI/dV$. Since transport through a single $\mathrm{C_{60}}$
molecule has been demonstrated\cite{Park} and the synthesis of endohedral
fullerenes is also feasible,\cite{Almeida} such an experiment is possible
with present-day apparatus. Besides the typical Coulomb-blockade behavior
we predict a characteristic fine structure of the Coulomb-blockade peaks in
$dI/dV$ due to the exchange coupling of the $\mathrm{C_{60}}$ spin to the
spin $3/2$ of the encapsulated nitrogen atom. It should be mentioned that
the discussion of transport through $\mathrm{P@C_{60}}$ proceeds quite
analogously and yields qualitatively identical results.

\section{Theory}

In our model the $\mathrm{N@C_{60}}$ molecule is treated as a quantum dot
and the leads, labeled as L (left) and R (right), as macroscopic charge
reservoirs. Relaxation in the leads is assumed to be sufficiently fast so
that the electron distributions in the leads can be described by Fermi
functions. As $\mathrm{C_{60}}$ generally prefers to be singly or doubly
negatively charged,\cite{Hettich,Pederson,YaL94,Wierzbowska,LPX04}
we assume that electronic
transport through the molecule involves only the threefold-degenerate LUMO
(lowest unoccupied molecular orbital).\cite{Laouini} Since we
concentrate on the fine structure of the differential conductance close to
degeneracy points, we assume that the fivefold-degenerate HOMO (highest
occupied molecular orbital), which lies about $7.5\,\mathrm{eV}$ below the
LUMO,\cite{Wierzbowska} remains fully occupied, whereas the threefold-degenerate
LUMO+1, about $1.7\,\mathrm{eV}$
above the LUMO,\cite{Wierzbowska} remains empty.\cite{rem.LUMOplus1} Charge
transfer from the nitrogen atom to the fullerene cage is assumed to be
negligible. When the LUMO is partially occupied, the net spin of the
electrons in the LUMO, $\mathbf{S}_{\mathrm{C_{60}}}$, couples to the spin
$3/2$ of the nitrogen atom, $\mathbf{S}_{\mathrm{N}}$, and the total spin
of the molecule is
$\mathbf{S}=\mathbf{S}_{\mathrm{C_{60}}}+\mathbf{S}_{\mathrm{N}}$. The
exchange interaction can be written in the simple form $-J\,
\mathbf{S}_{\mathrm{C_{60}}} \cdot \mathbf{S}_{\mathrm{N}}$ due to Hund's
first rule: Note that the three LUMO's and the three nitrogen
$p$ orbitals both have odd parity with a single nodal plane each, which can
be chosen as the $xy$, $yz$, or $zx$ plane. Consequently, there is only a
significant exchange interaction between any LUMO and the $p$ orbital of the
same symmetry. The exchange interaction can thus be written as a sum of
terms for the three pairs of orbitals. However, due to the strong
Hund's-rule coupling, the $p$ orbital spins combine to $S_{\mathrm{N}}=3/2$;
i.e., they
are all parallel. Then the spin of each $p$ orbital is
$\mathbf{S}_{\mathrm{N}}/3$, as
can be proved by projecting the spins onto the $S_{\mathrm{N}}=3/2$
subspace. Thus the
exchange terms can be combined to the simple scalar product. The full
Hamiltonian of the system then reads $H = H_{\text{d}} + H_{\text{leads}} +
H_{\text{t}}$, where
\begin{equation}
H_{\mathrm{d}} = \left( \varepsilon -e V_{\mathrm{g}} \right) n_{\text{d}}
  + \frac{U}{2}n_{\text{d}}(n_{\text{d}}-1)
  - J\, \mathbf{S}_{\mathrm{C_{60}}} \cdot \mathbf{S}_{\mathrm{N}}
\end{equation}
represents the molecular quantum dot,
\begin{equation}
H_{\text{leads}} = \sum_{\alpha=\text{L},\text{R}} \sum_{\mathbf{k} \sigma}
  \epsilon_{\alpha \mathbf{k}}
  a_{\alpha \mathbf{k} \sigma}^{\dagger} a_{\alpha \mathbf{k} \sigma}
\end{equation}
represents the leads, and
\begin{equation}\label{tunneling}
H_{\text{t}}= \sum_{\alpha=\text{L},\text{R}} \sum_{n \mathbf{k} \sigma}
  (t_{\alpha} a_{\alpha \mathbf{k} \sigma}^{\dagger} c_{n \sigma}
  + t_{\alpha}^{\ast} c_{n \sigma}^{\dagger} a_{\alpha \mathbf{k} \sigma})
\end{equation}
describes the tunneling between the dot and the leads, which is
assumed to be weak compared to typical excitation energies of the
molecule. Here, the
operator $c_{n \sigma}^{\dagger}$ creates an electron with spin $\sigma$ in
the molecular orbital $n$, while $a_{\alpha \mathbf{k} \sigma}^{\dagger}$
creates an electron in lead $\alpha$
with spin $\sigma$, momentum $\mathbf{k}$, and 
energy $\epsilon_{\alpha \mathbf{k}}$ relative to the Fermi energy.
$n_{\text{d}}=\sum_{n\sigma} c_{n\sigma}^\dagger c_{n\sigma}$
and $\textbf{S}_{\mathrm{C_{60}}}=\sum_{n\sigma\sigma'} c_{n\sigma}^\dagger
(\mbox{\boldmath$\sigma$}_{\sigma\sigma'}/2) c_{n\sigma'}$
are the number and spin operators of electrons on the dot, respectively.
Electron-electron interaction is taken into account by the
local Coulomb repulsion $U$ and the exchange interaction with the nitrogen
spin by the exchange coupling $J$.
The values of $\varepsilon$, $U$, and $J$ are not well known at present.
\textit{Ab initio} calculations\cite{Pederson,YaL94,Wierzbowska} indicate
that $\mathrm{{C_{60}}^{-}}$ is the ground state, whereas
$\mathrm{{C_{60}}^{2-}}$ has a slightly higher energy. This is in agreement
with the experimental observation of a very-long-lived metastable
$\mathrm{{C_{60}}^{2-}}$ (Ref.~\onlinecite{Hettich}). However, there are recent
contradicting \textit{ab initio} results predicting $\mathrm{{C_{60}}^{2-}}$
to be slightly bound relative to $\mathrm{{C_{60}}^{-}}$ (Ref.~\onlinecite{LPX04}).
For our numerical calculations we use $\varepsilon=-2.75\,\mathrm{eV}$ and 
$U=2.84\,\mathrm{eV}$ in accordance with Ref.~\onlinecite{YaL94}.
$J$ appears to be \emph{ferromagnetic}. We take
$J\sim 1\,\mathrm{meV}$ from \textit{ab initio} calculations of
Udvardi.\cite{Udvardi} This relatively strong exchange coupling is
consistent with the absence of electron-paramagnetic-resonance (EPR) signals
for $\mathrm{N@C}_{60}$ anions with charges $-1$ through $-5$, while the
signal is present for the neutral molecule and the hexa-anion.\cite{JGW00}
Note that the exchange coupling is significantly smaller than the energy of 
relevant vibrational modes. 
The oscillations of the molecule as a whole have an energy of the
order of $5\,\mathrm{meV}$.\cite{Park}
The oscillations of the nitrogen atom within
the $\mathrm{C_{60}}$ have an energy of $13\,\mathrm{meV}$,\cite{Uhlik}
whereas the modes of the $\mathrm{C_{60}}$ cage lie at much higher
energies.

We next derive rate equations for this model starting from the
equation of motion for the density matrix $\rho$,\cite{Blum,Mitra,Koch}
${d \rho_I(t)}/{dt}=-i \left[H_{\mathrm{t}I},\rho_I(t)\right]$.
Here, the index $I$ denotes the interaction representation with respect 
to $H_{\mathrm{t}}$. Integration and iteration gives\cite{Blum,Mitra}
\begin{equation}\label{Neumann2}
\begin{split}
\frac{d \rho_{I}(t)}{dt} = & -i \left[H_{\mathrm{t} I}(t),\rho_{I}(0)\right] \\
& {}- \int_{0}^{t} dt'\, \left[H_{\mathrm{t} I}(t),
  \left[H_{\mathrm{t} I}(t'),\rho_{I}(t')\right]\right].
\end{split}
\end{equation}
Assuming that the leads are weakly affected by the
quantum dot and neglecting 
correlations between the two, $\rho_{I}(t)$ can be replaced by the
direct product of the \emph{reduced density matrix} of the dot,
$\rho_{\text{d}I}(t) \equiv \text{Tr}_{\text{leads}}\, \rho_{I}(t)$,
and the density matrix $\rho_{\text{leads}}$ of the leads,
$\rho_{I}(t) \approx \rho_{\text{d}I}(t) \otimes \rho_{\text{leads}}$.
We then obtain
\begin{equation}\label{Neumann3}
\frac{d \rho_{\text{d}I}(t)}{dt} = - \! \int_{0}^{t}\! dt'\,
  \text{Tr}_{\text{leads}} \left[H_{\text{t}I}(t),
  \left[H_{\text{t}I}(t'),\rho_{\text{d}I}(t') \otimes
  \rho_{\text{leads}}\right]\right].
\end{equation}
Returning to the Schr\"odinger representation and
using the Markov approximation\cite{Blum,Mitra}
$\rho_{\text{d}I}(t') \approx \rho_{\text{d}I}(t)$, we find
\begin{equation}\label{Neumann4}
\begin{split}
& \frac{d \rho_{\text{d}}(t)}{dt} = -i\left[H_{\text{d}},\rho_{\text{d}}\right]
  - \text{Tr}_{\text{leads}} \int_{0}^{\infty} \!\! dt'\,
  \Big[H_{\text{t}}, \\ 
& \quad\left[ e^{-i(H_{\text{d}} + H_{\text{leads}})t'}
  H_{\text{t}} e^{i(H_{\text{d}} + H_{\text{leads}})t'} ,\rho_{\text{d}}(t)
  \otimes \rho_{\text{leads}} \right] \Big]
\end{split}
\end{equation}
as the equation of motion for $\rho_{\text{d}}$.
This expression shows that
the tunneling Hamiltonian $H_{\text{t}}$ is treated in second-order
perturbation theory. Taking the trace over
the degrees of freedom of the leads produces Fermi functions according to
\begin{equation}
\text{Tr}_{\text{leads}}\,
  \rho_{\text{leads}}\, a_{\alpha \mathbf{k} \sigma}^{\dagger}
  a_{\alpha' \mathbf{k}' \sigma'}
= \delta_{\alpha \alpha'} \delta_{\mathbf{k} \mathbf{k}'}
  \delta_{\sigma \sigma'}\, f(\epsilon_{\alpha \mathbf{k}}-\mu_{\alpha}),
\end{equation}
where $\mu_{\alpha}$ denotes the chemical potential of lead $\alpha$
due to the applied source-drain voltage $V$.
Expanding the nested commutators in Eq.~(\ref{Neumann4}) and inserting
Eq.~(\ref{tunneling}) gives eight terms:
\begin{equation}\label{Neumann5}
\begin{split}
& \frac{d \rho_{\text{d}}(t)}{dt} = - \int_{0}^{\infty}\!\! dt' \!
  \sum_{\alpha n n' \mathbf{k} \sigma} |t_{\alpha}|^2 \\
& {}\times \Big\{ f(\epsilon_{\alpha \mathbf{k}}-\mu_{\alpha})
  e^{i\epsilon_{\alpha \mathbf{k}} t'} c_{n \sigma} e^{-i H_{\text{d}} t'}
  c_{n' \sigma}^{\dagger} e^{i H_{\text{d}} t'} \rho_{\text{d}}(t) \\
& \;{}+ \left[ 1-f(\epsilon_{\alpha \mathbf{k}}-\mu_{\alpha}) \right]
  e^{-i \epsilon_{\alpha \mathbf{k}}  t'} c_{n \sigma}^{\dagger}
  e^{-i H_{\text{d}}  t'}
  c_{n' \sigma} e^{i H_{\text{d}} t'}\! \rho_{\text{d}}(t) \\
& \;{}- \left[ 1-f(\epsilon_{\alpha \mathbf{k}}-\mu_{\alpha}) \right]
  e^{i \epsilon_{\alpha \mathbf{k}}  t'} c_{n \sigma} \rho_{\text{d}}(t)
  e^{-i H_{\text{d}}  t'} c_{n' \sigma}^{\dagger} e^{i H_{\text{d}} t'} \\
& \;{}- f(\epsilon_{\alpha \mathbf{k}}-\mu_{\alpha})
  e^{-i\epsilon_{\alpha \mathbf{k}} t'} c_{n \sigma}^{\dagger}
  \rho_{\text{d}}(t) e^{-i H_{\text{d}}  t'} c_{n' \sigma}
  e^{i H_{\text{d}} t'} \\
& \;{}- \left[ 1-f(\epsilon_{\alpha \mathbf{k}}-\mu_{\alpha}) \right]
  e^{-i\epsilon_{\alpha \mathbf{k}} t'} e^{-i H_{\text{d}}  t'}
  c_{n \sigma} e^{i H_{\text{d}}  t'}\! \rho_{\text{d}}(t)
  c_{n' \sigma}^{\dagger} \\
& \;{}- f(\epsilon_{\alpha \mathbf{k}}-\mu_{\alpha})
  e^{i \epsilon_{\alpha \mathbf{k}}  t'} e^{-i H_{\text{d}} t'}
  c_{n \sigma}^{\dagger} e^{i H_{\text{d}}  t'} \rho_{\text{d}}(t)
  c_{n' \sigma} \\
& \;{}+ f(\epsilon_{\alpha \mathbf{k}}-\mu_{\alpha})
  e^{-i\epsilon_{\alpha \mathbf{k}} t'} \rho_{\text{d}}(t)
  e^{-i H_{\text{d}}  t'} c_{n \sigma} e^{i H_{\text{d}}  t'}
  c_{n' \sigma}^{\dagger} \\
& \;{}+ \left[ 1-f(\epsilon_{\alpha \mathbf{k}}-\mu_{\alpha}) \right]
  e^{i \epsilon_{\alpha \mathbf{k}}  t'} \rho_{\text{d}}(t)
  e^{-i H_{\text{d}} t'} c_{n \sigma}^{\dagger} e^{i H_{\text{d}} t'}\!
  c_{n' \sigma} \Big\}.
\end{split}
\end{equation}
The probability of the dot being in the many-particle state $|n\rangle$ is
$P^n \equiv \langle n | \rho_{\text{d}} (t) | n \rangle$.
Introducing the overlap matrix elements
$C_{mn}^{\sigma} \equiv \langle m | \sum_{i} c_{i \sigma} | n \rangle$ and
$C_{mn}^{\sigma \dagger} \equiv \langle m | \sum_{i} c_{i \sigma}^{\dagger}
| n \rangle$
and identifying the integrals in Eq.~(\ref{Neumann5}) as $\delta$ functions
we can write Eq.~(\ref{Neumann5}) as a set of rate equations
\begin{equation}\label{rateEq}
\frac{dP^n}{dt} = \sum_{m\neq n} P^{m} R_{m \rightarrow n}
  - P^{n} \sum_{m \neq n} R_{n \rightarrow m},
\end{equation}
with transition rates
\begin{equation}\label{rates}
\begin{split}
R_{n \rightarrow m} = & \sum_{\alpha \sigma}
  2\pi\,|t_{\alpha}|^2\, D_{\alpha}\,
  f(\epsilon^{\text{d}}_{m}-\epsilon^{\text{d}}_n-\mu_{\alpha}) \\
& {}\times \left( |C_{nm}^{\sigma}|^2 + |C_{mn}^{\sigma}|^2 \right).
\end{split}
\end{equation}
Here, $\epsilon^{\text{d}}_n$ is the energy of the many-particle state
$|n\rangle$ of the dot and $D_{\alpha}$ denotes the density of states 
per spin species in
lead $\alpha$, which we take to be constant and equal for both leads.
The matrix elements 
$C_{mn}^{\sigma}$ ($C_{mn}^{\sigma \dagger}$) can only be finite
if the electron number of state $| n \rangle$ 
is larger (smaller) by 1 than the electron number of state $| m \rangle$.
We are interested in the \emph{stationary state}, which corresponds to
$dP^n/dt=0$ for all states $|n\rangle$.

In deriving Eq.~(\ref{rateEq}) we have assumed that the density matrix
$\rho_{\text{d}}$ is \emph{completely diagonal}. This assumption requires
some thought since many of the eigenstates of our molecular quantum dot are
degenerate so that one might expect finite off-diagonal components even in
the stationary state. However, this is not the case: Let $U$ be a unitary
matrix that leaves the dot Hamiltonian $H_{\text{d}}$ invariant. With any
stationary density matrix $\rho_{\text{d}}$, $U\rho_{\text{d}}U^\dagger$ is
another solution. Now suppose that there exists a stationary solution
$\rho_{\text{d}}$ that is \emph{not} diagonal within a block of degenerate
states. Then one can choose $U$ so that $U\rho_{\text{d}}U^\dagger$ is
diagonal since the nonzero off-diagonal components have been assumed to
connect degenerate states (we exclude the case of accidental degeneracy).
But then $U\rho_{\text{d}}U^\dagger$ has \emph{unequal} diagonal
components---i.e., probabilities $P^n$---for symmetry-related states. This is
clearly unphysical. On the other hand, if $\rho_{\text{d}}$ is already
diagonal with degenerate dot states having equal diagonal components, any
allowed transformation $U$ leaves $\rho_{\text{d}}$ invariant.

The current operator for lead $\alpha=\mathrm{L},\mathrm{R}$ reads\cite{Mahan}
\begin{equation}
I_{\alpha} = i \left[H,N_\alpha \right]
  = -i \sum_{n \mathbf{k} \sigma} (t_{\alpha} c_{n \sigma}^{\dagger}
  a_{\alpha \mathbf{k} \sigma} - t_{\alpha}^{\ast}
  a_{\alpha \mathbf{k} \sigma}^{\dagger} c_{n \sigma}) ,
\end{equation}
where $N_\alpha$ is the number operator for electrons in lead $\alpha$.
Tracing out the leads we arrive at an expression for the expectation
value of the current:
\begin{equation}\label{current}
\begin{split}
\langle I_{\alpha} \rangle = &~2 \pi D_{\alpha} |t_{\alpha}|^2
  \sum_{m l \sigma} \Big( f(\epsilon^{\text{d}}_{l} - 
\epsilon^{\text{d}}_{m} -\mu_{\alpha}) |C_{ml}^{\sigma}|^2 \\
& - \left[ 1-f(\epsilon^{\text{d}}_{m} - \epsilon^{\text{d}}_{l}-\mu_{\alpha}) 
\right] |C_{lm}^{\sigma}|^2 \Big) P^m. \\
\end{split}
\end{equation}
We here consider the symmetric case $t_{\text{L}}=t_{\text{R}}$ and
$\mu_{\text{L}}=-\mu_{\text{R}}=V/2$.

As there are ${6 \choose i}$ possible ways of filling the threefold-degenerate
$\mathrm{C_{60}}$ LUMO with $i$ electrons according to the 
Pauli principle and as the nitrogen atom possesses a spin $3/2$, solving
the rate equations and calculating the current
reduces to an eigenvalue problem  of dimension $4\times \sum_{i=0}^{6}
{6 \choose i} = 256$.

\section{Results and discussion}

\begin{figure}[t]
\begin{center}
\psfrag{V}[][][1][0]{$V$ (V)}
\psfrag{Vsd}[][][1][0]{\scriptsize{$V$ (V)}}
\psfrag{I}[][][1][0]{$I$ ($2 \pi e D_{\text{L}} t_{\text{L}}^2/\hbar$)}
\psfrag{n}[][][1][0]{$\langle n \rangle$}
\includegraphics[width=3.2in,clip]{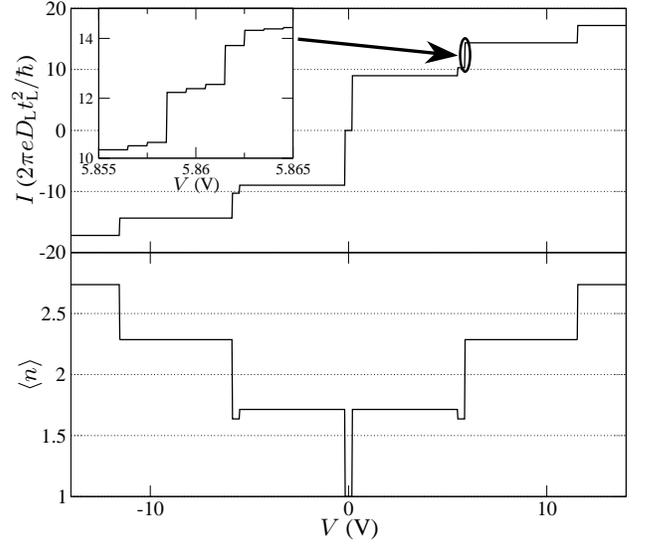}
\caption{Current $I$ and average occupation $\langle n \rangle$ of the
$\mathrm{C_{60}}$ LUMO as  a function of the source-drain voltage 
$V\equiv\mu_{\text{L}}-\mu_{\text{R}}$ for
$\varepsilon=-2.75\,\mathrm{eV}$, $U=2.84\,\mathrm{eV}$, $J=1\,\mathrm{meV}$,
$V_g=0\,\mathrm{V}$, and $T=0.01\,\mathrm{K}$. The inset shows the
fine structure of one particular Coulomb-blockade step.}\label{fig_current}
\end{center}
\end{figure}

The \textit{I-V} characteristics plotted in Fig.~\ref{fig_current} show a
conductance gap for $|V|<0.18\,\mathrm{V}$ due to Coulomb blockade.
Our numerical results show that  the current $I$ is symmetric with respect
to the applied source-drain voltage $V$ in accordance with the high
symmetry of the fullerene molecule. Each step in the main \textit{I-V} curve
corresponds to the opening of additional current channels. Simultaneously,
the average occupation $\langle n \rangle$ of the dot changes. For the
parameters chosen above, the $\mathrm{{C_{60}}^{-}}$ state is the
ground state.\cite{YaL94} At the first step, the potential drop becomes
large enough to allow transitions between the nearly degenerate charge
states $-1$ and $-2$, as the chemical potential $\mu_{\text{L}}=V/2$
reaches the value assumed for the ionization energy
$E(\mathrm{{C_{60}}^{2-}}) -
E(\mathrm{{C_{60}}^{-}})=\varepsilon+U=0.09\,\mathrm{eV}$. At the second
step, transitions between the charge states $-1$ and $0$ become
possible, etc.

Our results for the occupation probabilities reveal that detailed balance
is satisfied for the broad  plateaus in Fig.~\ref{fig_current}---i.e.,
$P^{n} R_{n \rightarrow m} = P^{m} R_{m \rightarrow n}$. As a consequence,
the dot occupation probabilities $P^n$ for all \emph{occupied} states are
identical in the limit $T\rightarrow0$, as the transition rates $R_{n
\rightarrow m}$ are symmetric for each pair $n$, $m$ of occupied states.
This also accounts for the fact that the average occupation $\langle n
\rangle$ is exactly unity for $V=0\,\mathrm{V}$, increases to $(24
\times 1+60 \times 2)/(24+60)=12/7\approx 1.71$ at the first step, when the
molecule is in one of 24 singly charged or 60 doubly charged states with
equal probability, and decreases to $(4 \times 0 + 24 \times 1 + 60 \times
2)/(4+24+60)= 18/11 \approx 1.64$ at the second step, when 4 additional
neutral states become available, etc. Furthermore, we find that each
Coulomb-blockade step shows a characteristic fine structure, which we
discuss below.

\begin{figure}[thb]
\begin{center}
\psfrag{Vg}[][][1][0]{$V_{\mathrm{g}}$~(mV)}
\psfrag{V}[t][][1][0]{$V$~(mV)}
\psfrag{a}[b][][1][0]{\footnotesize{\textbf{(a)}}}
\psfrag{b}[b][][1][0]{\footnotesize{\textbf{(b)}}}
\psfrag{c}[b][][1][0]{\footnotesize{\textbf{(c)}}}
\psfrag{d}[b][][1][0]{\footnotesize{\textbf{(d)}}}
\psfrag{x1}[][][1][0]{\tiny{$88$}}
\psfrag{x2}[][][1][0]{\tiny{$90$}}
\psfrag{x3}[][][1][0]{\tiny{$92$}}
\psfrag{x4}[][][1][0]{\tiny{$2928$}}
\psfrag{x5}[][][1][0]{\tiny{$2930$}}
\psfrag{x6}[][][1][0]{\tiny{$2932$}}
\psfrag{x7}[][][1][0]{}
\psfrag{y1}[][][1][0]{\tiny{$-5$}}
\psfrag{y2}[][][1][0]{\tiny{$0$}}
\psfrag{y3}[][][1][0]{\tiny{$5$}}
\psfrag{y4}[][][1][0]{}
\includegraphics[width=3.41in,angle=0]{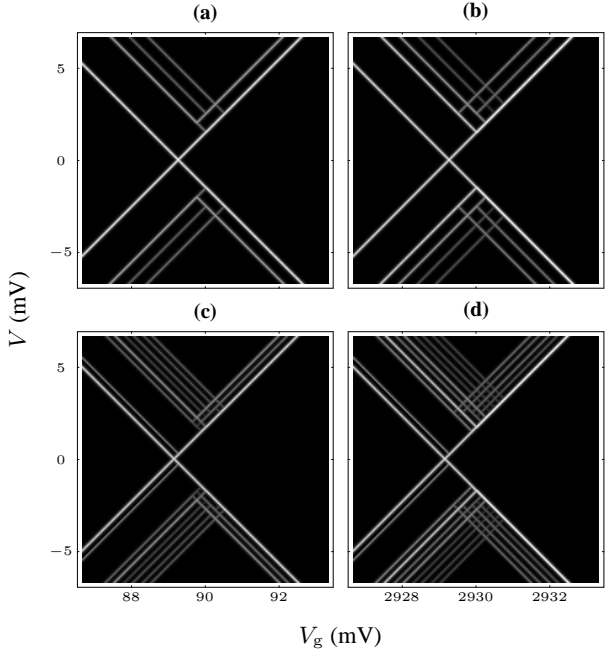}
\caption{Gray-scale plots of the differential conductance $dI/dV$ as a
function of source-drain voltage $V$ and gate voltage
$V_{\mathrm{g}}$ for $T=0.1\,\mathrm{K}$. Shown are two particular ranges of
gate voltages close to the degeneracy points between charge states $-1$ and
$-2$ (a),(c) and between $-2$ and $-3$ (b),(d).
(a) and (b) show results for vanishing magnetic field and
(c) and (d) for $B=2\,\mathrm{T}$.}\label{fig_diamonds}
\end{center}
\end{figure}

The calculation of the differential conductance $dI/dV$ as a function of 
source-drain voltage $V$ and gate voltage $V_{\mathrm{g}}$ shows the usual
\emph{Coulomb diamonds}; see Fig.~\ref{fig_diamonds}. Close to the
degeneracy points between different charge states we observe a relatively
complex fine structure, corresponding to the steps in the inset of
Fig.~\ref{fig_current}. We assume very low temperatures, $k_BT\ll J$, to
exhibit the structure more clearly. At higher temperatures the peaks in
$dI/dV$ are thermally broadened. In the following, we briefly explain the
physics behind the fine structure, taking Fig.~\ref{fig_diamonds}(a) as an
example.

\begin{figure}[tb]
\vspace*{0.15in}
\begin{center}
\tiny
\psfrag{a}[b][][1][0]{\footnotesize{\textbf{(a)}}}
\psfrag{b}[b][][1][0]{\footnotesize{\textbf{(b)}}}
\psfrag{n1}{$N=1$}
\psfrag{n2}{$N=2$}
\psfrag{DeltaE}{$\Delta E_{\mathrm{exc}}~\mathrm{(meV)}$}
\psfrag{DeltaE1}{$+0.75$}
\psfrag{DeltaE2}{$-0.75$}
\psfrag{DeltaE3}{$-1.75$}
\psfrag{DeltaE4}{$+0.25$}
\psfrag{DeltaE5}{$-1.25$}
\psfrag{DeltaE6}{$+1.25$}
\psfrag{DeltaE7}{$-0.75$}
\psfrag{DeltaE8}{$+0.75$}
\psfrag{DeltaE9}{$-1.25$}
\psfrag{DeltaE10}{$-0.25$}
\psfrag{DeltaE11}{$+1.25$}
\psfrag{DeltaE12}{$+1.75$}
\psfrag{S1}{$S_{\mathrm{C_{60}}}=1/2$,~$S=1$}
\psfrag{S2}{$S_{\mathrm{C_{60}}}=1/2$,~$S=2$}
\psfrag{S3}{$S_{\mathrm{C_{60}}}=1$,~$S=1/2$}
\psfrag{S4}{$S_{\mathrm{C_{60}}}=1$,~$S=3/2$}
\psfrag{S5}{$S_{\mathrm{C_{60}}}=0$,~$S=3/2$}
\psfrag{S6}{$S_{\mathrm{C_{60}}}=1$,~$S=5/2$}
\hspace{-0.7in}\includegraphics[height=1.5in,width=2.6in,angle=0]{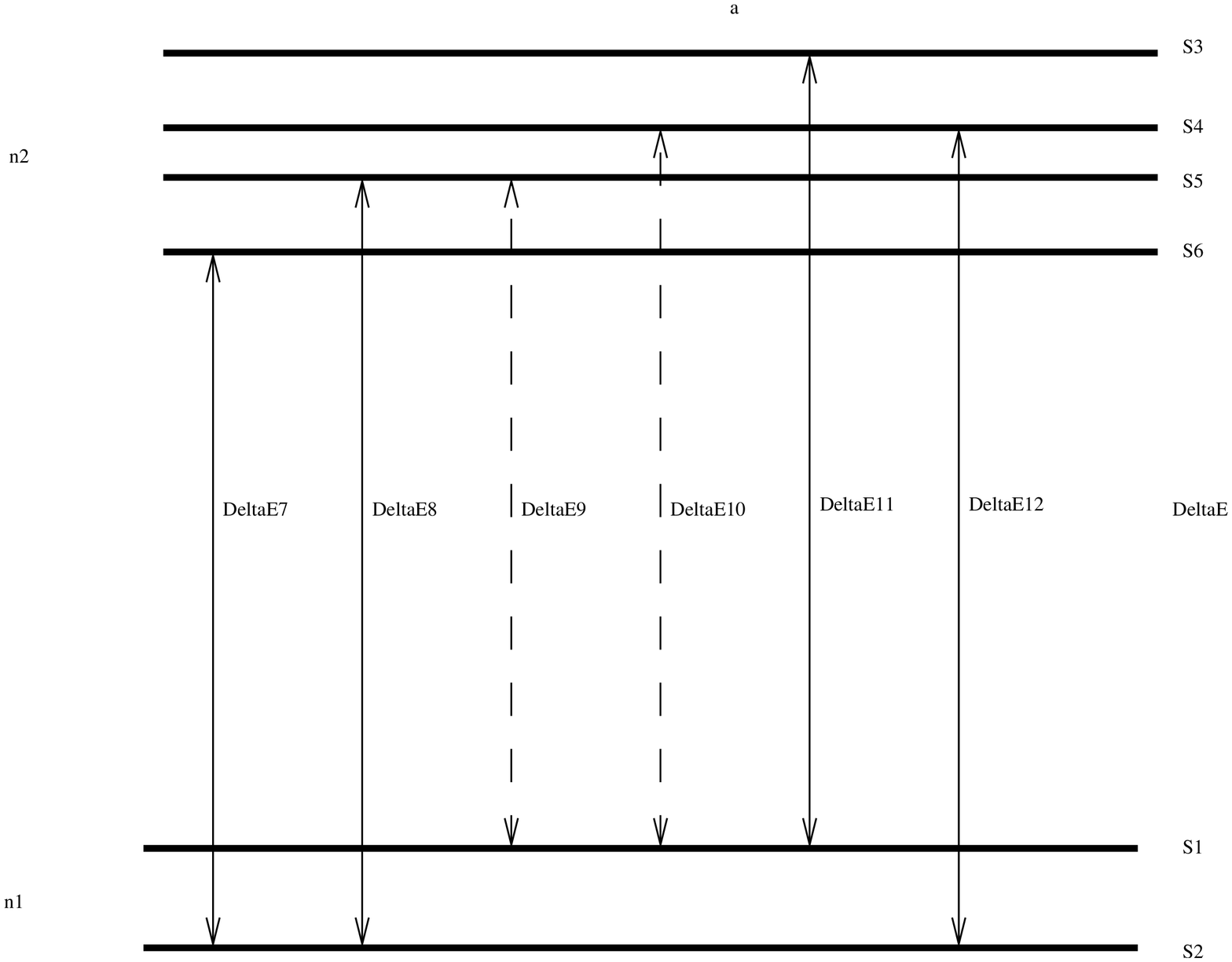}\\
\vspace{0.4in}
\hspace{-0.7in}\includegraphics[height=1.5in,width=2.6in,angle=0]{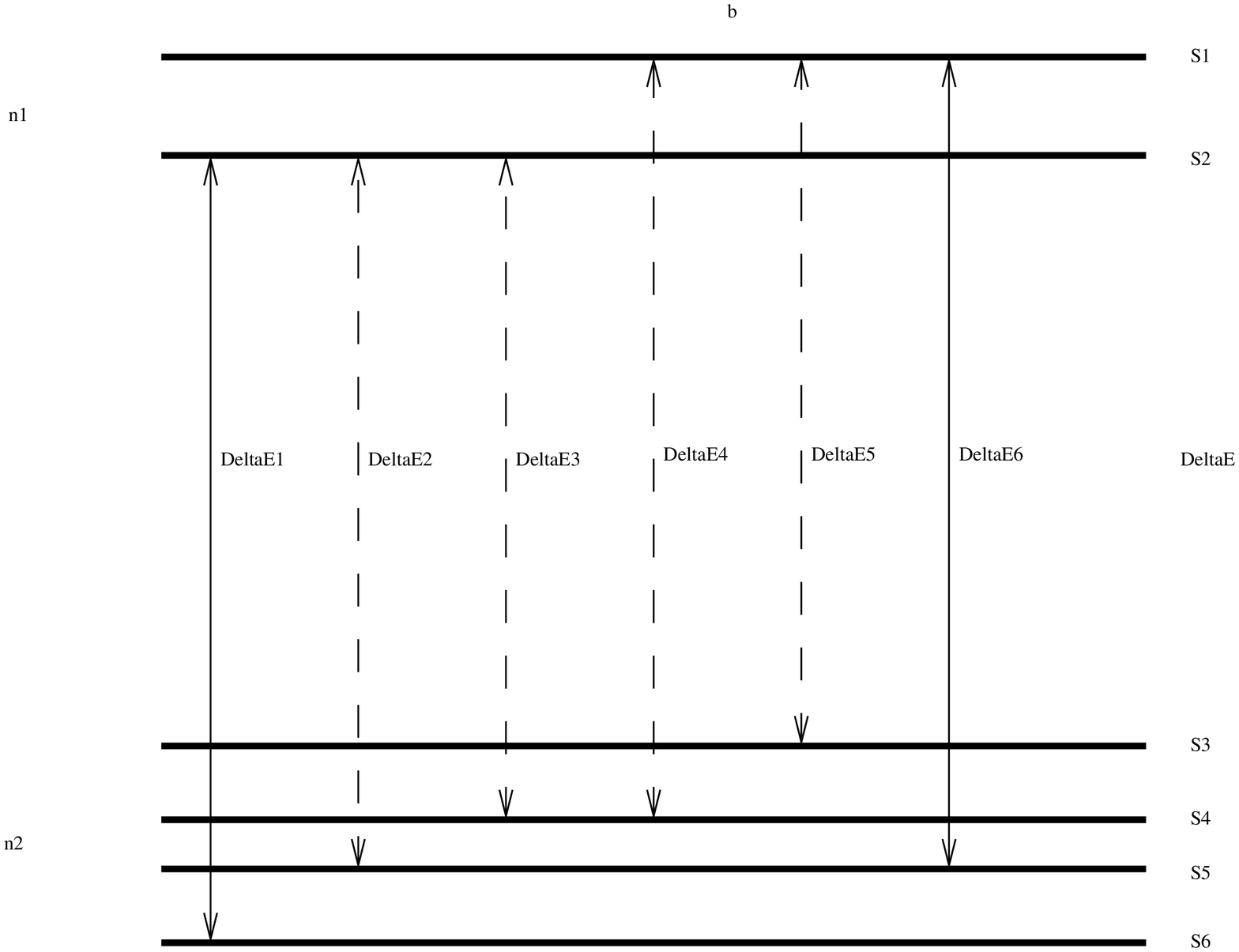}
\caption{Energy levels and all allowed transitions between
many-particle states with one ($N=1$) and
two ($N=2$) electrons, taking into account spin
excitations. (a)~Situation with the $N=1$ multiplet
lower in energy than the $N=2$ multiplet.
(b)~Reverse case.}\label{fig_levels}
\end{center}
\end{figure}

Since the $\mathrm{C_{60}}$ spin $S_\mathrm{C_{60}}$, the spin of the
nitrogen atom, $S_\mathrm{N}$, and the total spin $S$ (where
$S=|S_\mathrm{C_{60}}-S_\mathrm{N}|,...,S_\mathrm{C_{60}}+S_\mathrm{N}$)
are good quantum numbers, the exchange energy is
\begin{equation}
E_{\mathrm{exc}} = - \frac{J}{2} \big[ S(S+1)
  - S_{\mathrm{C_{60}}}(S_{\mathrm{C_{60}}}+1)
  - S_{\mathrm{N}}(S_{\mathrm{N}}+1) \big],
\end{equation}
which leads to the level splitting illustrated in Fig.~\ref{fig_levels}.
The structure in Fig.~\ref{fig_diamonds}(a) arises from transitions
between charge states $-1$ and $-2$, taking spin excitations into account.
In equilibrium ($V=0\,\mathrm{V}$) only the ground
state of the dot is occupied, which is the $\mathrm{{C_{60}}^{-}}$ state with
$S_{\mathrm{C_{60}}}=1/2$ and $S=2$ for $V_{\mathrm{g}}$ smaller than 
the degeneracy point $V_{\mathrm{g}}^0$ and the $\mathrm{{C_{60}}^{2-}}$ state 
with $S_{\mathrm{C_{60}}}=1$ and $S=5/2$ for
$V_{\mathrm{g}}>V_{\mathrm{g}}^0$; cf.~Fig.~\ref{fig_levels}(a). 
For $V_{\mathrm{g}}<V_{\mathrm{g}}^0$
the \emph{first} peak in $dI/dV$ at nonzero $V$
originates from the transition with $S_{\mathrm{C_{60}}}=1/2
\rightarrow 1$ and $S=2 \rightarrow 5/2$, corresponding to a gain of
exchange energy of $\Delta E_{\mathrm{exc}}=-0.75\,\mathrm{meV}$. The
\emph{second} peak results from a transition with
$S_{\mathrm{C_{60}}}= 1/2 \rightarrow 0$, $S=2 \rightarrow 3/2$, and $\Delta
E_{\mathrm{exc}}=+0.75\,\mathrm{meV}$. Simultaneously, the
transitions with $S_{\mathrm{C_{60}}}=1/2 \rightarrow 0$, $S=1 \rightarrow
3/2$, and $\Delta E_{\mathrm{exc}}=-1.25\,\mathrm{meV}$ and
$S_{\mathrm{C_{60}}}= 1/2 \rightarrow 1$, $S=1 \rightarrow 3/2$, and $\Delta
E_{\mathrm{exc}}=-0.25\,\mathrm{meV}$ are enabled [dashed lines in
Fig.~\ref{fig_levels}(a)].  Although energetically possible, these
transitions are not excited at lower source-drain voltages, because the
lower levels are unoccupied. The last two peaks  belong to transitions
with $S_{\mathrm{C_{60}}}= 1/2 \rightarrow 1$, $S=1 \rightarrow 1/2$, and
$\Delta E_{\mathrm{exc}}=+1.25\,\mathrm{meV}$ and $S=1/2 \rightarrow 1$,
$S=2 \rightarrow 3/2$, and $\Delta E_{\mathrm{exc}}=+1.75\,\mathrm{meV}$.
Note that the values of $\Delta E_{\mathrm{exc}}$ account for the
level spacing.

The situation is different for $V_\mathrm{g}$ significantly larger than
$V_{\mathrm{g}}^0$, where we observe only two peaks;
cf.~Fig.~\ref{fig_diamonds}(a). As soon as the transition from the
$\mathrm{{C_{60}}^{2-}}$ ground state into the lowest $\mathrm{{C_{60}}^{-}}$
state with $S_{\mathrm{C_{60}}}=1 \rightarrow 1/2$, $S=5/2 \rightarrow 2$,
and $\Delta E_{\mathrm{exc}}=+0.75\,\mathrm{meV}$ becomes possible, the
transitions corresponding to $\Delta E_{\mathrm{exc}}=-0.75\,\mathrm{meV}$,
$-1.75\,\mathrm{meV}$, $+0.25\,\mathrm{meV}$, and $-1.25\,\mathrm{meV}$
[dashed lines in Fig.~\ref{fig_levels}(b)] are also enabled. Again the
latter four would be energetically possible at lower $V$, but do not appear
as peaks of $dI/dV$, since the corresponding lower levels are unoccupied.
In the vicinity of $V_{\mathrm{g}}^0$ we find that the slope of several
lines abruptly changes sign. This corresponds to the situation where two
levels connected in Fig.~\ref{fig_levels} by a transition cross as
$V_{\mathrm{g}}$ is varied. The fine structure in
Fig.~\ref{fig_diamonds}(b) can be discussed analogously. The structure is
different for all degeneracy points and can thus serve as a
\emph{fingerprint} of the particular charge transition. This should be
useful since the zero of the $V_{\mathrm{g}}$ axis is often shifted
significantly from one experiment to the next.

Selection rules for single-electron tunneling require that $\Delta
S_{\mathrm{C_{60}}} = \pm 1/2$ and $\Delta S = \pm 1/2$. The different
brightness of the peaks in Fig.~\ref{fig_diamonds} is correlated with the
number of transitions that are possible at a given source-drain voltage.
Each allowed transition may be thought of as one current channel.

Experimentally, the \emph{magnetic} origin of the fine structure is most
conclusively tested by observing the behavior in a magnetic field. For
ionized $\mathrm{C_{60}}$ in lattices and in solution, the orbital moment is
quenched.\cite{KKO91,TMG96} We assume that the fields generated by the
electrodes in a break junction are also sufficiently strong to quench the
orbital moment. Then the molecule couples to a magnetic induction $B$ only
through the \emph{spin} moments, described by the new Hamiltonian
\begin{equation}\label{Zeeman}
H' = H - g\mu_B B S_{\mathrm{C_{60}}}^z
  - g\mu_B B S_{\mathrm{N}}^z
  = H - g\mu_B B S^z .
\end{equation}
Here, $\mu_B$ is the Bohr magneton and $g$ is the $g$ factor, which is
$g\approx 2$ for both the nitrogen spin $\mathbf{S}_{\mathrm{N}}$ and the
$\mathrm{C_{60}}$ spin.
We choose a many-particle basis of simultaneous eigenstates of
$n_{\text{d}}$, $S_{\mathrm{C_{60}}}$, $S$, and $S^z$. Then the only difference is
that additional Zeeman energies appear in our expression for the transition
rates, Eq.~(\ref{rates}). In Figs.~\ref{fig_diamonds}(c) and \ref{fig_diamonds}(d)
we show $dI/dV$
for the same parameters as in Figs.~\ref{fig_diamonds}(a) and \ref{fig_diamonds}(b) 
but with
$B=2\,\mathrm{T}$. As expected, the peaks split, but in addition 
several peaks are absent since they are not allowed by the selection
rules. For example, for $V_{\mathrm{g}}>V_{\mathrm{g}}^0$ the first peak is
due to a transition with $S_{\mathrm{C_{60}}} = 1 \rightarrow 1/2$, $S =
5/2 \rightarrow 2$, and $N=2\rightarrow 1$. Since the initial state has all
spins aligned in parallel, one electron tunneling out of the dot can only
reduce $S^z$ so that there is only a \emph{single} peak in
$dI/dV$.

To summarize, we have presented a theory for transport through a single
$\mathrm{N@C_{60}}$ molecule weakly coupled to metallic electrodes. Our
results for the differential conductance $dI/dV$ as a function of the
source-drain and gate voltages show Coulomb-blockade and exhibit a
characteristic fine structure of the Coulomb-blockade peaks due to the
coupling of the $\mathrm{C_{60}}$ spin to the spin of the encapsulated
nitrogen atom.

\acknowledgments

We would like to thank W. Harneit, J. Koch, A. \mbox{Mitra}, and F. von Oppen for
helpful discussions and the Deutsche Forschungsgemeinschaft for support
through Sonderforschungsbereich 290.


\begin{thebibliography}{99}
\bibitem{rem.LUMOplus1}Since the LUMO+1 has even parity, whereas the
nitrogen $p$ orbitals are odd, the exchange interaction with the LUMO+1 is
negligible.
\end{thebibliography}
\end{document}